\documentclass[12pt]{article}
\begin{document}
\begin{center}
\bf{A Comment on the Invariance of the Speed of Light}\\

Harihar Behera\\

Patapur, P.O.: Endal, Jajpur-755023, Orissa, India\\

E-mail : harihar@iopb.res.in
\end{center}

\begin{abstract}
The invariance of the speed of light in all inertial frames is shown to be 
an inevitable consequence of the relativity principle of special relativity 
contrary to the view held by Hsu and Hsu in taiji relativity where the speed 
of light is no longer a universal constant.The present approach is not 
only new but also much simpler than the existing approaches.\\
\end{abstract}
\vskip 0.5cm

\newpage 
The two postulates of special relativity, viz.,$ (i)$ the relativity principle 
and $ (ii) $ the postulate on the speed of light, are the basics of a theory that has been tremendously successful in describing a wide range of phenomena, 
although there exist derivations of Lorentz transformations (LT) without 
the postulate on the speed of light [1-8]. However, using the relativity 
principle only, Hsu and Hsu [9] have developed a general theory, named taiji 
relativity, which has four dimensional symmetry and is consistent with 
experiments although the speed of light is no longer a universal constant 
within its framework. In the development of taiji relativity the authors 
have considered the usual case of two inertial frames $  F $ and $ F' $ and 
to simplify the discussion defined the speed of light to be constant and 
isotropic in the frame $ F $  by synchronizing the clocks in that frame according 
to the usual method of special relativity. The authors have made this 
definition in one frame only and could just as easily have used the $ F
$ frame, as all frames are equivalent, so this procedure does not select a 
preferred frame. As the principle of relativity itself does not specify how 
the $ F $ and $ F' $ clocks should be related, the theory does not tell us 
anything about the speed of light in any other frame, they say [9-12]. But 
here in this communication, the invariance of the speed of light in all inertial frames is shown to be a natural consequence of the relativity principle 
contrary to the view held by the authors [9-12].\\

Consider two inertial frames $ F $ and $ F' $  which are in uniform relative 
motion. An event may be characterized by specifying the co-ordinates 
 $ (x,y,z,t) $ of the event in $ F $ and the event is characterized by the co-ordinates $ (x', y', z', t') $ in $ F' $. Let us proceed to find a transformation between $ ( x,y,z,t) $ and $ ( x', y', z', t') $ without the light-speed postulate of special relativity. To  simplify the algebra let the relative velocity$ v  $ of $ F $ and $ F' $ be along a common $ x/x' $ axis with corresponding planes parallel. Also at  the instant the origins $ 0 $ and $ 0' $ coincide, we let the clocks there  to read $ t = 0 $ and $ t'= 0 $ respectively. The homogeneity of space and time  in inertial frames requires that the transformations must be linear so that the simplest form they can take is\\                        \begin{equation} x' = k (x - vt);\>\>\>   y' = y;\>\>\>   z' = z;\>\>\>  t' = lx + mt                                                                        \end{equation} \\
In order to determine the values of the three co-efficients $ k,l,$ and $ m $ we shall  use the idea of the homogeneity and isotropy of space. Let us assume that at  the time $ t = 0 $ a spherical electromagnetic wave left the origin of $ F $  which  coincides with the origin of $ F' $ at that moment. Let the speed of 
light be assumed as $ c $ and $ c' $ in $ F $ and $ F' $ respectively with 
$ c  \neq  c' $. We know that the two frames $ F $ and $ F' $ are  equivalent. Therefore the electromagnetic wave propagates in all directions  in each inertial frame as the space is homogeneous and isotropic. By the  equivalence of inertial frames (i.e. relativity principle), the wave forms  (fronts) of the electromagnetic wave in the two frames $ F $ and $ F' $ must  be equivalent or similar, otherwise it would be possible to determine from  the shape of the wave front which one i.e.$  F $ or $ F' $ is at rest or in  uniform motion from which the absolute nature of motion or rest may be  inferred contrary to the relativity principle. Again the wave front of a  spherical electromagnetic wave in an isotropic and homogeneous medium  describes a sphere whose radius expands with time at a rate equal to its speed in that medium. Thus the equation of the spherical wave fronts in the  two inertial frames must take the forms:
\begin{equation} x^{2} + y^{2} + z^{2} = c^{2}t^{2}                             \end{equation}                                                                 \begin{equation}{x'}^{2} +{y'}^{2} +{z'}^{2} ={c'}^{2}{t'}^ {2}                         \end{equation}                                                                 when the transformations (1) are substituted in (3), we get 
\begin{equation} 
(k^{2} - l^{2}c'^{2})x^{2} + y^{2} + z^{2} -2(lmc'^{2} - k^{2}v) =
(m^{2}c'^{2} - k^{2}v^{2})t^{2}  
\end{equation}                                           
In order for the expression (4) to agree with (2), which represents the same 
thing, we must have
\begin{eqnarray} 
(i) && k^{2} - l^{2}c'^{2} = 1, \nonumber \\
(ii) && m^{2}c'^{2} - k^{2}v^{2} = c^{2},\nonumber \\ 
(iii) && lmc'^{2} - k^{2}v = 0 
\end{eqnarray}       

This set of three equations (5)  when solved for $ k,l $ and $ m $ yield
\begin{eqnarray} 
(i) && k = (1 - {v^{2}}/{c^{2}})^{-1/2} ,\nonumber \\
(ii) && l = -(c/c')(v/c^{2})(1 - {v^{2}}/{c^{2}})^{-1/2} ,\nonumber \\
(iii) && m = (c/c')(1 - {v^{2}}/{c^{2}})^{-1/2}                                \end{eqnarray}  
Now in view of these values of $k, l$ and $ m $, the transformations
(1) can be represented in the following matrix form, viz.,     
\begin{equation}\left( \begin{array}{c} x'\\ y'\\ z'\\ t'
    \end{array}\right) = A \left( \begin{array}{c} x \\ y \\ z \\ t
    \end{array}\right) ,\>\>\>\> {\rm where} \   A = \left(
    \begin{array}{cccc} k & 0 & 0 & -kv \\ 0 & 1 & 0 & 0 \\ 0 & 0 & 1
      & 0 \\ -kv/cc' & 0 & 0 & kc/c' \end{array} \right)      
\end{equation}

The inverse transformations for $ x, y, z, t $, then given by
\begin{equation}\left( \begin{array}{c} x \\ y \\ z \\ t
\end{array}\right) = A^{-1} \left( \begin{array}{c} x' \\  y' \\ z'\\
  t' \end{array}\right)  ,\>\>\>\> {\rm where} \  A^{-1} = \left(
\begin{array}{cccc} k  &  0  & 0&  kvc'/c  \\  0  &  1  &  0  &  0  \\
  0  &  0  &  1 & 0  \\  kv/c^{2}  &  0  &  0 & kc'/c  \end{array}
\right)  
\end{equation} 
Hence the transformation for $ x $ becomes
\begin{equation} x = k (x' + vc't'/c)
\end{equation}
But relativity principle demands that the transformation for $ x $ must be given  by [13]:
\begin{equation} x = k(x' + vt')
\end{equation}
Thus Eq. (9) contradicts Eq. (10). Such a contradiction arises because of 
our assumption that $ c  \neq  c' $. Hence in the order for Eq. (9) to be 
in accord with the relativity principle (so with Eq. (10)),we must have $ c = c'  $.

From the above discussion, we may conclude that the postulation on the speed 
of light in special relativity is an inevitable consequence of the 
relativity principle taken in conjunction with the idea of the homogeneity 
and isotropy of space and the homogeneity of time in all inertial frames. 
The present approach is physically distinct from and logically simpler than 
the standard special relativity, as we make use of only one postulate. The 
idea of constructing a relativity theory by using only the relativity 
principle has also been discussed by Ritz, Tolman and Pauli [14], but the 
present approach is not only new but also much simpler than the existing approaches and has been given as a preliminary report in [15].\\

The author thanks Prof. N.Barik, Dept. of Phys. Utkal University, 
Bhubaneswar, P.C.Naik, Dept. of Physics, D.D.College, Keonjhar and N.K. Behera, Dept. of Chem. U.N. College, Soro for fruitful 
discussions and suggestions.The author also acknoledges the help
received from the Institute of Physics,Bhubaneswar for using its Library and 
Computer Centre for this work.


\end{document}